\documentclass[a4paper,10pt]{iopart}
\usepackage{iopams}
\usepackage{graphicx}
\usepackage{cite}
\usepackage[colorlinks=true,linkcolor=blue,citecolor=blue]{hyperref}

\newcommand{\Ip}{I_\mathrm{p}}
\newcommand{\Iref}{I_\mathrm{ref}}
\newcommand{\qref}{q_\mathrm{ref}}
\newcommand{\nref}{n_\mathrm{ref}}
\newcommand{\mrm}[1]{\mathrm{#1}}
\newcommand{\wA}{\omega_\mathrm{A}}
\newcommand{\VA}{V_\mathrm{A}}
\newcommand{\helena}{\texttt{HELENA}}
\newcommand{\castork}{\texttt{CASTOR-K}}
\newcommand{\mishka}{\texttt{MISHKA}}
\newcommand{\alfven}{Alfv\'{e}n}
\newcommand{\aefamily}[3]{^{#1}\!E_{#2,#3}}
\newcommand{\doublepicture}[2]{
    \includegraphics[width=0.480\textwidth]{#1}
    \hspace{0.025\textwidth}
    \includegraphics[width=0.480\textwidth]{#2}}

\begin{document}

\title
[Sensitivity of alpha-particle--driven \alfven{} eigenmodes \ldots]
{Sensitivity of alpha-particle--driven \alfven{} eigenmodes to q-profile
variation in ITER scenarios}

\date{\today}

\author{P. Rodrigues$^1$, A. C. A. Figueiredo$^1$,
D. Borba$^1$,
R. Coelho$^1$,
L. Fazendeiro$^1$,
J. Ferreira$^1$,
N. F. Loureiro$^{1,2}$,
F. Nabais$^1$,
S. D. Pinches$^3$,
A. R. Polevoi$^3$,
and S. E. Sharapov$^4$}

\address{$^1$Instituto de Plasmas e Fus\~{a}o Nuclear, Instituto Superior
T\'{e}cnico, Universidade de Lisboa, 1049-001 Lisboa, Portugal.}

\address{$^2$Plasma Science and Fusion Center, Massachusetts Institute of
Technology, Cambridge, MA 02139, USA.}

\address{$^3$ITER Organization, Route de Vinon-sur-Verdon, 13067 St
Paul-lez-Durance Cedex, France.}

\address{$^4$CCFE, Culham Science Centre, Abingdon OX14 3DB, United Kingdom.}

\ead{par@ipfn.ist.utl.pt}

\begin{abstract}
A perturbative hybrid ideal-MHD/drift-kinetic approach to assess the stability
of alpha-particle--driven \alfven{} eigenmodes in burning plasmas is used to
show that certain foreseen ITER scenarios, namely the $\Ip = 15$~MA baseline
scenario with very low and broad core magnetic shear, are sensitive to small
changes in the background magnetic equilibrium. Slight variations (of the order
of $1\%$) of the safety-factor value on axis are seen to cause large changes in
the growth rate, toroidal mode number, and radial location of the most unstable
eigenmodes found. The observed sensitivity is shown to proceed from the very low
magnetic shear values attained throughout the plasma core, raising issues about
reliable predictions of alpha-particle transport in burning plasmas.
\end{abstract}

\maketitle

\section{Introduction}

Plasma heating during the burning regime in tokamak reactors will rely upon the
energy of fusion-born alpha-particles, which must be kept confined to ensure
efficient heating and that power fluxes remain within the design values of the
ITER plasma facing components~\cite{fasoli.2007}. However, such particles can
drive \alfven{} Eigenmodes (AEs) unstable and be thus transported away from the
plasma core, which would hamper the burning
process~\cite{breizman.2011,gorelenkov.2014}. In order to predict the level of
alpha-particle redistribution and loss that is expected for a given fusion
scenario, the most unstable AEs need to be identified first so that their
stability properties can be employed in further analysis.

Understanding the complex interplay between energetic suprathermal particles and
AEs is a key step in the fusion research
effort~\cite{breizman.2011,sharapov.2013,gorelenkov.2014}, particularly in
preparation for future burning-plasma experiments. Recent research concerning
ITER~\cite{pinches.2015,lauber.2015,rodrigues.2015,figueiredo.2016} has been
focusing on the $15$~MA baseline scenario, with on-axis safety factor close to
unity and low magnetic shear throughout the plasma core~\cite{polevoi.2002}. Low
magnetic-shear profiles are indeed expected to take place during planned ITER
operation due to sawtooth activity, which periodically redistributes the
toroidal current density within a large mixing region that may extend to about
half of the plasma minor radius because safety-factor values at the boundary are
low ($q_\mrm{b} \sim 3$)~\cite{porcelli.1996,hender.2007}. In addition, low
magnetic shear was found to play a significant role in the nonlinear
stabilization of microturbulence by suprathermal pressure
gradients~\cite{citrin.2013} --- an important observation especially in cases
where sheared toroidal momentum is insufficient to provide the said
stabilization~\cite{burrell.1997,doyle.2007}.

In this work, a perturbative hybrid ideal-MHD/drift-kinetic plasma model is used
to find how the stability properties of AEs change in response to small
variations of the background magnetic equilibrium. Of particular interest are
the net growth rate, wave number, and frequency of the most unstable AEs. These
properties are shown to be significantly affected by small changes of the
safety-factor profile, which are achieved through slight variations of the total
plasma current. Such high sensitivity is also shown to be caused by the
low levels of magnetic shear present in the scenario under analysis.

\section{Particle-wave interaction model}

Routine stability assessments in burning plasmas can be accomplished with an
hybrid MHD--drift-kinetic model of particle-wave
interaction~\cite{rodrigues.2015}. Here, ideal-MHD theory is used to describe
thermal species (DT fuel ions, electrons, He ash and other impurities), whose
energy distribution functions are assumed to be local Maxwellians. The radial
dependence of their temperature and particle-number density is an input to the
model. A similar input must also be provided for the density of the
suprathermal, diluted, fusion-born alpha-particle population, which is assumed
to be isotropic in pitch angle. Its energy distribution function is described by
the model~\cite{pinches.1998}
\begin{equation}
    f_\mrm{sd}(E) = \frac{C_\mrm{N}}{E^{3/2} + E^{3/2}_\mrm{c}}
        \: \mrm{erfc} \bigl[ (E - E_{0})/\Delta_\mrm{E} \bigr],
\label{eq:slowing.down}
\end{equation}
where $C_\mrm{N}$ is a normalization constant while $E_\mrm{c}$,
$\Delta_\mrm{E}$, and $E_0$ are radius-independent parameters, and $\mrm{erfc}$
is the complementary error function.

The response of the non-Maxwellian alpha-particle population to an ideal-MHD
perturbation of the thermal plasma is found solving the linearized drift-kinetic
equation~\cite{porcelli.1994}, valid in the limit
\begin{equation}
    \omega \big/ \Omega_\alpha \sim \bigl( k_\perp \rho_\alpha \bigr)^2 \ll 1,
\label{eq:drift.kinetic}
\end{equation}
with $\omega$ and $k_\perp$ the AE frequency and perpendicular wave number,
whereas $\Omega_\alpha$ and $\rho_\alpha$ are the alpha-particle gyro-frequency
and gyro-radius. This response gives rise to a small complex correction
$\delta\omega$ to the frequency $\omega$ of marginally stable AEs and the
alpha-particle contribution to their growth rate is then $\gamma_\alpha =
\mrm{Im}(\delta\omega)$. A similar procedure for each thermal species $j$
produces the corresponding Landau-damping contribution $\gamma_j$ to the
wave-particle energy exchange. Disregarding effects not contained in the
ideal-MHD framework (e.g., \alfven{}-continuum damping, radiative damping),
which cannot be modeled by the perturbative approach just described, the overall
AE growth rate is thus $\gamma_\alpha + \textstyle\sum_{j} \gamma_j$.

The workflow to assess the stability of a given plasma configuration is as
follows~\cite{rodrigues.2015}: a magnetic equilibrium is first computed with
\helena{}~\cite{huysmans.1991}, using pressure and current-density profiles
obtained from transport modelling, and then all possible AEs are found by
intensively scanning over a frequency and wave-number range with the ideal-MHD
code \mishka{}~\cite{mikhailovskii.1997}, while the energy transfer between them
and all plasma species is evaluated with the drift-kinetic code
\castork{}~\cite{borba.1999,nabais.2015}. The computational efficiency of the
\mishka{}/\castork{} pair is the key to handle the very large number of AEs
involved in such systematic stability assessments.

\section{The reference case}

\begin{figure*}
\doublepicture{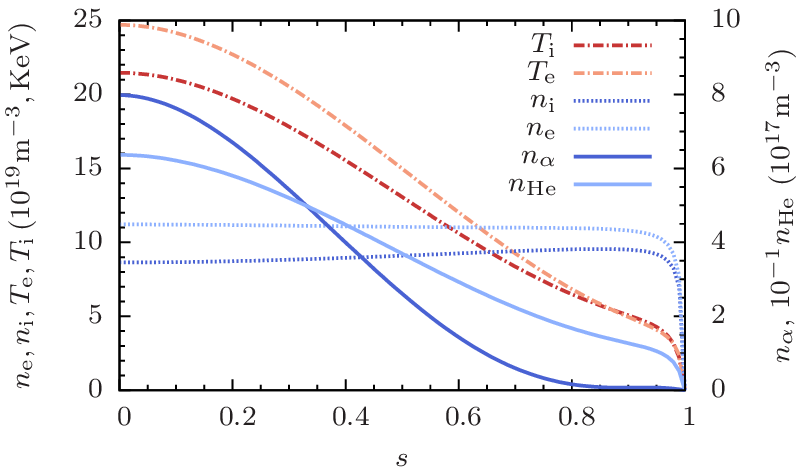}{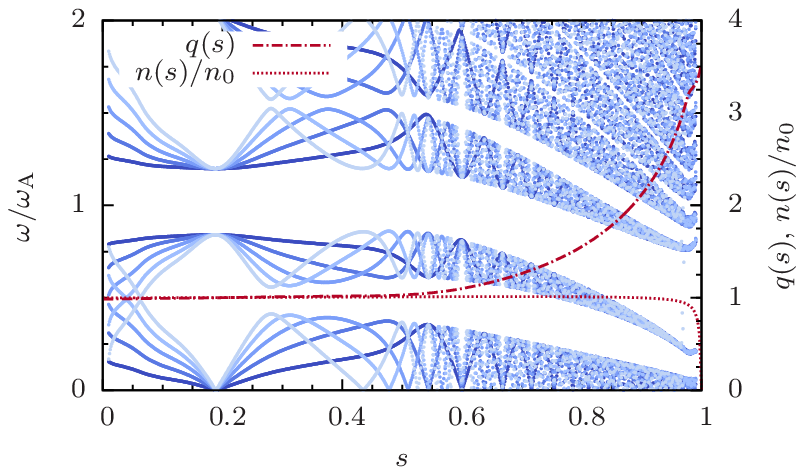}
\caption{\label{fig:reference.scenario}
Radial distribution of the plasma-species densities and temperatures (left).
Ideal \alfven{} continua for toroidal mode numbers $n = 10,\ldots,50$ (from dark
to light hues), safety factor, and normalized mass density (right).}
\end{figure*}

The radial distribution of each species' particle-density and temperature for
the ITER $\Ip = 15$~MA baseline scenario~\cite{polevoi.2002} were found with the
transport code \texttt{ASTRA}~\cite{pereverzev.2008} and are displayed in
\fref{fig:reference.scenario}, where $s^2 = \psi \big/ \psi_\mrm{b}$, $\psi$ is
the poloidal flux, and $\psi_\mrm{b}$ is its value at the boundary.  Other
relevant parameters are the on-axis magnetic field $B_0 = 5.3$~T, the minor
radius $a = 2$~m, and the magnetic-axis location at $R_0 = 6.4$~m (not to be
confused with the tokamak's geometric axis $R_\mrm{vac} = 6.2$~m). The DT fuel
mix ratio is $n_\mrm{D} / n_\mrm{T} = 1$ and their combined density is
$n_\mrm{i} = n_\mrm{D} + n_\mrm{T}$. Peaked temperature profiles contrast with
DT-ion and electron density distributions, which are flat almost up to the
plasma edge. In turn, fusion alpha-particles are mostly concentrated in the
core, with an almost constant gradient $\rmd n_\alpha/\rmd s$ for $0.3 \lesssim
s \lesssim 0.5$.

Flat mass-density distributions up to the plasma edge, like the one plotted in
\fref{fig:reference.scenario}, contribute to the closing of the frequency gaps
in the \alfven{} continuum arising from the coupling between distinct poloidal
harmonics. Consequently, AEs with frequencies in such gaps can hardly extend
towards the plasma boundary without interacting with the \alfven{} continuum and
thus undergo significant damping. This property acts as a filter regarding the
type of AEs that can be found for the particular plasma state being considered.
Actually, the safety-factor profile also depicted in
\fref{fig:reference.scenario} (right panel) is almost flat in the core region
($s \lesssim 0.5$), yielding well separated gaps for toroidal mode numbers $n
\gtrsim 10$. Highly localized low-shear toroidicity-induced AEs (LSTAEs), with
only two dominant poloidal harmonics, are therefore expected to arise in the
core. Conversely, on the outer half of the plasma the magnetic shear is higher,
radial gap separation is smaller, and AEs become broader, encompassing a large
number of poloidal harmonics and extending to the edge. In so doing, they
interact with the \alfven{} continuum and are thus excluded from further
analysis, which will be dominated by $n \gtrsim 10$ highly localized LSTAEs.

Contrary to the safety-factor profile, which establishes each AE location, the
square root of the density profile influences only the value of the \alfven{}
velocity. Therefore, no significant sensitivity is expected to small changes in
$n_\mrm{i}$ or $n_\mrm{e}$. On the other hand, and for the same value on-axis,
density distributions less flat than that in \fref{fig:reference.scenario} have
lower density near the edge, which contributes to raise the local continuum
frequency and therefore allow broader AEs to extend up to the edge. Without
significant interaction with the \alfven{} continuum, these broader AEs could
influence the overall stability properties of the considered scenario. A
detailed study of the consequences of such density-profile shaping is, however,
beyond the scope of this work.

The $(\omega,\mathbf{k})$-space scan carried out by \mishka{} finds the radial
structure of all AEs with toroidal number $n$ in the range $1 \leqslant n
\leqslant 50$ and poloidal harmonics $n - 1 \leqslant m \leqslant n + 15$. The
limit of 17 poloidal harmonics is set by pragmatic considerations: more
harmonics would benefit the convergence of broad-width AEs only, which would
nonetheless undergo continuum damping at some radial location since their
frequency gap is closed. In turn, the upper limit for $n$ is set by the
drift-kinetic ordering in~\eref{eq:drift.kinetic} as
\begin{equation}
    k_\perp \rho_\alpha \lesssim 1, \quad \mathrm{whence} \quad
        n \lesssim \bigl( s / q \bigr) \big/ \bigl( \rho_\alpha / a \bigr)
            \approx 50,
\label{eq:kperp.ralpha.1}
\end{equation}
with $\rho_\alpha / a \approx10^{-2}$ the normalized alpha-particle gyro-radius,
$k_\perp \sim \bigl( n q \bigr) \big/ \bigl( a s \bigr)$, $q \approx 1$, and $s
\approx 0.5$. For each $n$, the frequency range $0 \leqslant \omega /
\omega_\mrm{A} \leqslant 1$ [where $\wA = \VA / R_0$, with $\VA$ the on-axis
\alfven{} velocity] is sampled in small steps of size $2 \times 10^{-5}$. Next,
\castork{} evaluates the energy exchange between every AE found and each of the
plasma species considered, yielding the corresponding growth (or damping) rate.
The parameters of the energy distribution-function model
in~\eref{eq:slowing.down} for the fusion-born alpha particles were taken at a
radial location ($s \approx 0.4)$ near the maximum gradient of their density
profile, yielding the values $E_\mrm{c} = 730$~KeV, $\Delta_\mrm{E} = 50$~KeV,
and $E_0 = 3.5$~MeV.

\begin{figure*}
\doublepicture{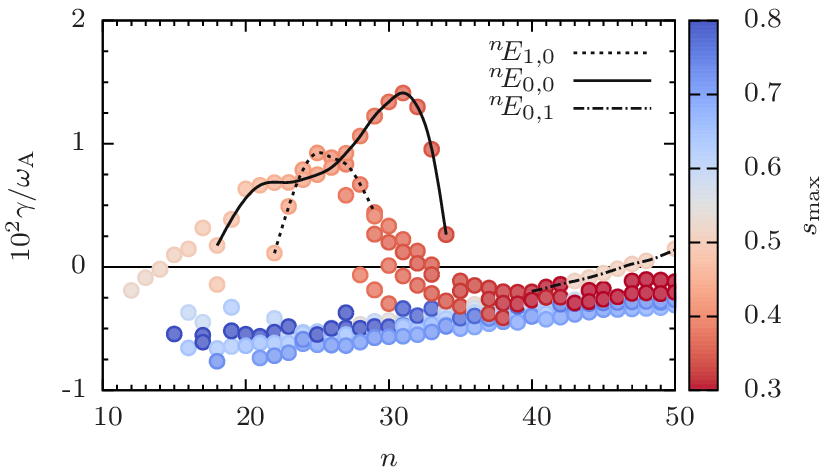}{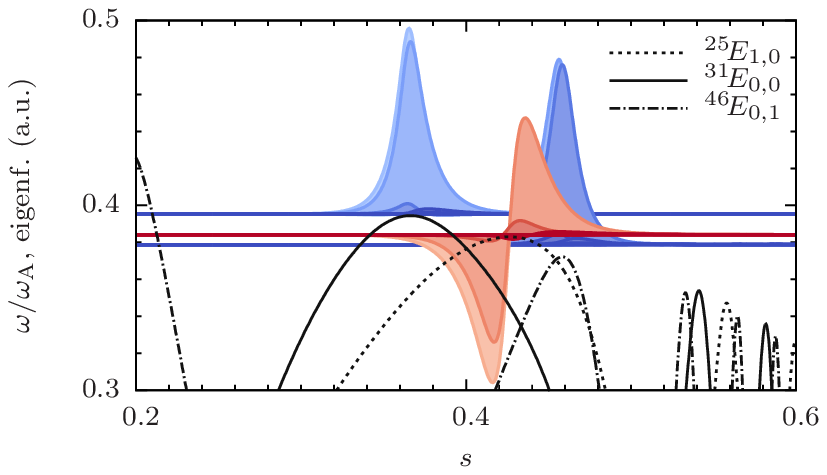}
\caption{\label{fig:stability.reference}
Distribution of the normalized growth rate $\gamma/\wA$ by $n$ for $\Iref$, with
each TAE colored by the radial location of its maximum amplitude, and three TAE
families identified by dotted, solid, and dash-dotted lines (left). Example
eigenfunctions of the AE families $\aefamily{n}{0}{1}$, $\aefamily{n}{1}{0}$,
and $\aefamily{n}{0}{0}$ (right), with baselines at their normalized frequencies
(filled curves) and their corresponding ideal \alfven{} continua (solid, dotted,
and dash-dotted lines).}
\end{figure*}

The stability assessment is summarized in \fref{fig:stability.reference} (left
panel) for the reference scenario where the plasma current $\Ip$ takes the
reference value $\Iref = 15$~MA and the on-axis safety factor is $\qref =
0.987$. This is essentially a subset of previous results~\cite{rodrigues.2015},
here restricted to toroidicity-induced AEs (TAEs) to make the presentation of
results clearer (whence the upper limit $\omega \big/ \wA \leqslant 1$ set on
the AEs' frequency). The most unstable TAEs found have $20 \lesssim n \lesssim
30$ and lie in the core ($0.3 \lesssim s_\mrm{max} \lesssim 0.5$, with
$s_\mrm{max}$ the location of their maximum amplitude), where $\rmd
n_\alpha/\rmd s$ is highest and the magnetic shear is lowest. Conversely, AEs
located in the outer half of the plasma ($0.5 \lesssim s_\mrm{max} \lesssim
0.8$) are mostly stable due to smaller values of the alpha-particle density
$n_\alpha$ and its gradient $\rmd n_\alpha/\rmd s$.

Three lines are plotted in \fref{fig:stability.reference} (left) connecting TAEs
belonging to three families that will play a key role in the ensuing discussion.
These families are denoted as $\aefamily{n}{l}{p}$, meaning that their members
are LSTAEs with even ($E$) parity and $l$ zeros, with $p$ being the difference
between the first dominant poloidal harmonic and the toroidal number $n$. A
member of each family has its radial structure depicted in the right panel of
\fref{fig:stability.reference}, where they can be identified by their respective
\alfven{} continua.

\section{Small changes in magnetic equilibria}

The reference value $\Iref = 15$~MA considered in the stability analysis that
led to \fref{fig:stability.reference} is only nominal. In practice, ITER
operation under this baseline scenario will exhibit values of the plasma current
which are close, but of course not rigorously equal to, the reference $\Iref$.
Therefore, details of the safety-factor profile near the magnetic axis are not
accurately established and can, eventually, change the stability properties of
the AEs found. Other factors that may cause a similar impact include the
presence of pressure anisotropy concurrently with low magnetic-shear
configurations~\cite{hole.2013}.

\begin{figure*}
\doublepicture{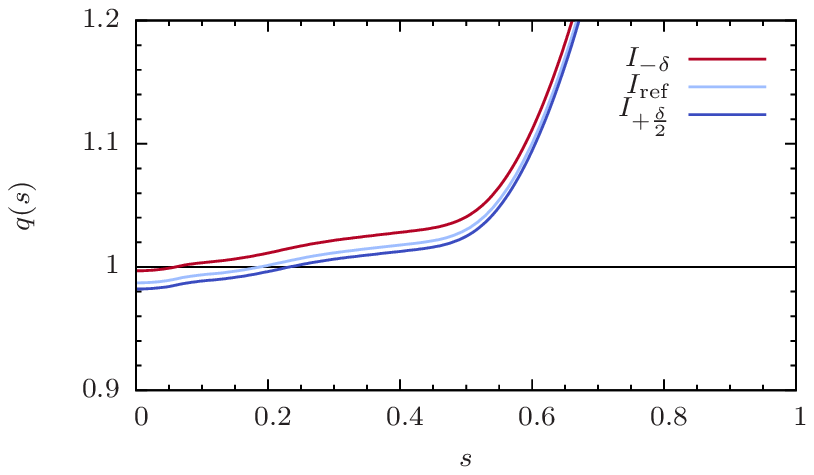}{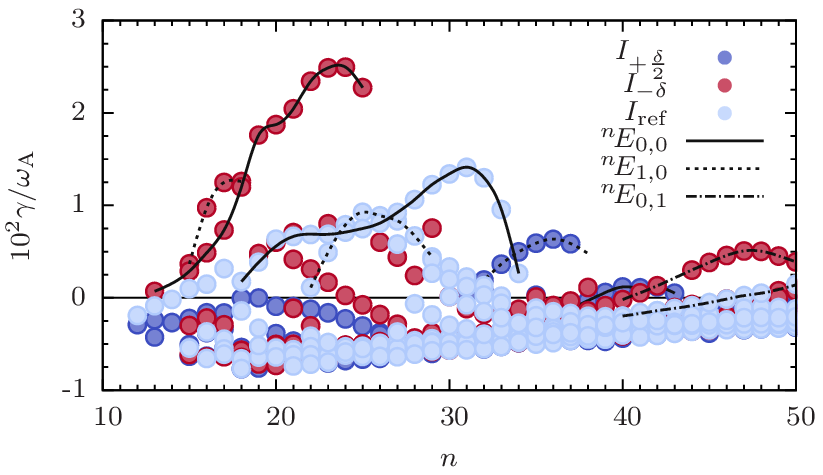}
\caption{\label{fig:sensitivity.1}
Left: Safety-factor profiles for the reference plasma current and two slightly
different $\Ip$ values. Right: Distribution of the normalized growth rate
$\gamma/\wA$ by $n$ for the three plasma-current values, with indication of the
three TAE families.}
\end{figure*}

To explore the dependence of AE stability properties on safety-factor
uncertainty, two different magnetic equilibria are next considered, in addition
to the reference one discussed in the previous Section. These equilibria are
obtained by changing $\Ip$ from $\Iref$ by the small amounts $-\delta$ and
$\delta/2$, with $\delta = 0.16$~MA, whilst keeping the same equilibrium
profiles $p'(\psi)$ and $f(\psi) f'(\psi)$. The resulting safety-factor profiles
are plotted in \fref{fig:sensitivity.1} (left) with the reference one for
comparison. As expected, the on-axis safety factor value $q_0$ changes only
slightly by circa $1\%$ and $0.5\%$ respectively, thus following the magnitude
of $\Ip$ variations away from $\Iref$. Moreover, the safety-factor slope in the
plasma core is kept almost unchanged in the two cases, with $q_0' \approx 0.07$.

The consequences to the stability properties are displayed in
\fref{fig:sensitivity.1} (right), where small variations ($\sim 1\%$) in $\Ip$
or $q_0$ are seen to cause large changes in the toroidal number ($\sim 20\%$)
and normalized growth rate ($\sim 50\%$) of the most unstable AEs. Raising
$q_0$ (and thus decreasing $\Ip$) pushes the most unstable AE families
($\aefamily{n}{0}{0}$ and $\aefamily{n}{1}{0}$) towards lower $n$ and up to
larger growth rates. A slight decrease in $q_0$ yields the opposite. In both
cases, the most unstable AEs are still even LSTAEs.

This extreme sensitivity to small changes in the background magnetic equilibrium
can be understood with the aid of the three conditions:
\begin{eqnarray}
    q \simeq q_0 + q_0' s,\label{eq:linear.q}%
    \label{eq:linear.q.condition}\\
    q = 1 + \frac{1}{2n},\label{eq:lstae.q}
    \label{eq:lstae.condition}
\end{eqnarray}
and
\begin{equation}
        k_\perp \Delta_\mrm{orb} \simeq \biggl( \frac{n q}{a s} \biggr)
            \biggr( \frac{a q}{\varepsilon \tilde{\Omega}} \biggr)
                \sim 1.
\label{eq:efficiency.condition}
\end{equation}
The first one is a linear representation of the safety factor profile in the
low-shear core and the second defines the resonant surface of each TAE in the
$\aefamily{n}{l}{0}$ families. In turn, the third relation is a condition for
efficient drive, with $\Delta_\mrm{orb} \sim \bigl( a q \bigr) \big/ \bigl(
\epsilon \tilde{\Omega} \bigr)$ the orbit width of an alpha-particle travelling
at the \alfven{} velocity with a very small pitch angle, $\tilde{\Omega} =
\Omega_\alpha / \wA$ its normalized gyro-frequency, and $\varepsilon = a / R_0$
the equilibrium inverse aspect ratio. Together, these three equations set the
three variables $n$, $q$, and $s$ (respectively, toroidal mode number, safety
factor, and radial location) corresponding to the most unstable AEs in terms of
the four parameters $q_0$, $q_0'$, $\varepsilon$, and $\tilde{\Omega}$.
Eliminating $s$ and $q$ from equations~\eref{eq:linear.q.condition},
\eref{eq:lstae.condition}, and~\eref{eq:efficiency.condition}, one finds the
toroidal number to follow the relation
\begin{equation}
    n + \frac{1 - 2 \zeta}{4n} + 1 = \zeta (1 - q_0),
\label{eq:unstable.n}
\end{equation}
which is written in terms of the dimensionless number
\begin{equation}
    \zeta \equiv \frac{\varepsilon \tilde{\Omega}}{q_0'} = \biggl(
        \frac{q}{q_0'} \biggr) \bigg/ \biggl( \frac{\Delta_\mrm{orb}}{a} \biggr).
\label{eq:zeta}
\end{equation}
Subtracting from equation~\eref{eq:unstable.n} its evaluation with the values
$\nref$ and $\qref$ corresponding to the reference case, gives
\begin{equation}
    \Biggl( 1 + \frac{2 \zeta - 1}{4 \nref \, n} \Biggr)
        \bigl( n - \nref \bigr) = - \zeta \bigl( q_0  - \qref\bigr),
\label{eq:sensitivity}
\end{equation}
which relates a variation of the on-axis safety factor with a corresponding
change in the toroidal number of the most unstable LSTAEs.

For the ITER scenario under consideration, parameters are $q_0' \approx 0.07$,
$\varepsilon \approx 0.3$, and $\tilde{\Omega} \approx 230$, whence $\zeta
\approx 10^3$. On the other hand, $n \sim \nref \sim 30$ and therefore $(2 \zeta
- 1) / (4 \nref n) \sim 1/2$. Because the prefactor in the left-hand side of
equation~\eref{eq:sensitivity} is of the order of unity, it is the large value
attained by $\zeta$ in the right-hand side that forces small changes of the
on-axis safety factor to cause large variations $n - \nref$. Also, one easily
checks that increasing $q_0$ above $\qref$ lowers $n$ below $\nref$ and
conversely, as observed in \fref{fig:sensitivity.1}. Moreover, the conditions in
equations~\eref{eq:linear.q},~\eref{eq:lstae.q},
and~\eref{eq:efficiency.condition} relate the radial location $s$ of the most
unstable AE with its toroidal number as
\begin{equation}
    \varepsilon \tilde{\Omega} s = n \biggl( 1 + \frac{1}{2 n} \biggr)^2,
\label{eq:radial.n}
\end{equation}
which predicts its displacement towards the core as $q_0$ increases and $n$
drops according to equation~\eref{eq:sensitivity}. As this happens, the AE
growth rate rises due to the larger number of alpha-particles found as it moves
inwards within the small region $0.3 \lesssim s \lesssim 0.4$, where $\rmd
n_\alpha / \rmd s$ is negative and almost constant
(\fref{fig:reference.scenario}). The consequences of decreasing $q_0$ (or
raising $\Ip$) are likewise explained.

The contribution of the alpha-particle population to the AEs growth rate
is~\cite{porcelli.1994}
\begin{equation}
    \gamma_\alpha \propto \omega  \frac{\partial f_\alpha}{\partial E}
            - n \frac{\partial f_\alpha}{\partial P_\phi},
\label{eq:alpha.drive}
\end{equation}
where $f_\alpha(E,P_\phi)$ is the unperturbed distribution function and
\begin{equation}
    P_\phi = \frac{\psi_\mrm{b}}{B_0 R_0^2} s^2
        + \frac{1}{\tilde{\Omega}} \frac{R B_{(\phi)}}{R_0 B}
            \frac{v_\parallel}{\VA}
\label{eq:p.phi}
\end{equation}
is the normalized toroidal canonical momentum of a particle moving with velocity
$v_\parallel$ parallel to a magnetic field with toroidal component $B_{(\phi)}$
and magnitude $B$. Because the derivative of~\eref{eq:slowing.down} with respect
to $E$ is always negative, alpha-particle drive results from the $P_\phi$
gradient. Such gradient relates with the radial derivative as
\begin{equation}
    \frac{2\psi_\mrm{b}}{B_0 R_0^2} \frac{\partial f_\alpha}{\partial P_\phi}
        \approx \frac{1}{s}\frac{\partial f_\alpha}{\partial s},
\label{eq:dPphi.ds}
\end{equation}
if the location $s$ is not so close to the magnetic axis so that terms of order
$(v_\parallel / \VA) \big/ (\varepsilon \tilde{\Omega})$ can be neglected.
Replacing equations~\eref{eq:efficiency.condition} and~\eref{eq:dPphi.ds}
in~\eref{eq:alpha.drive} and discarding the energy-gradient term, the AEs radial
location cancels out and one obtains
\begin{equation}
    \gamma_\alpha \propto
        - \frac{\varepsilon \tilde{\Omega}}{q^2}
            \frac{\partial f_\alpha}{\partial s}.
\label{eq:alpha.drive.optimal}
\end{equation}
Therefore, it may be asked if pushing unstable AEs further into the core, and
consequently out of the region where the gradient $\rmd n_\alpha / \rmd s$ is
strongest, may reduce $\gamma_\alpha$ and thus result in their stabilization.
Such inward push is achieved by slightly increasing $q_0$ (hence reducing
$\Ip$), which forces $n$ in equation~\eref{eq:sensitivity} and $s$ in
equation~\eref{eq:radial.n} to drop their values.

\begin{figure*}
\doublepicture{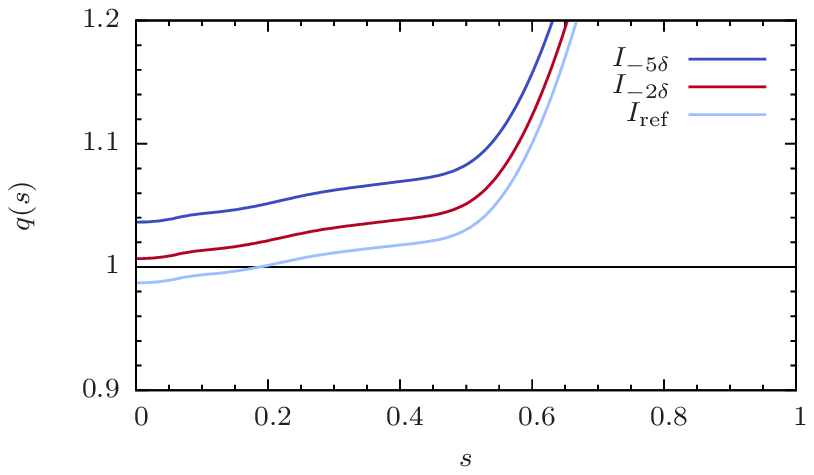}{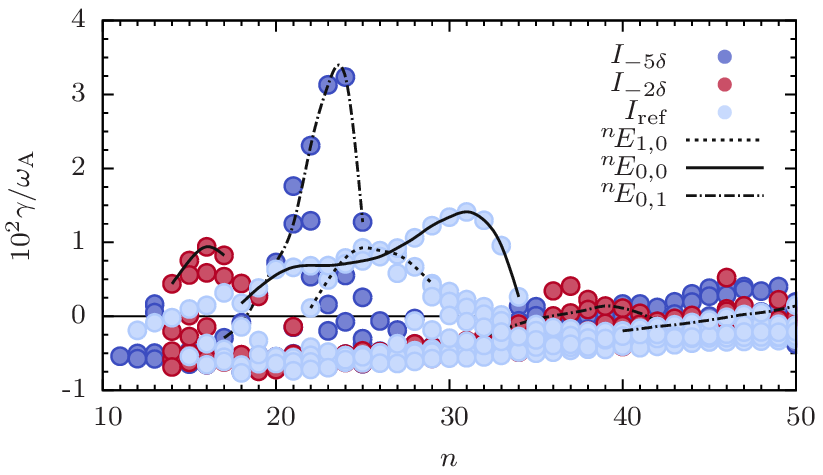}
\caption{\label{fig:sensitivity.2}
Safety-factor profiles for the reference plasma current and two other $\Ip$
values (left). Distribution of the normalized growth rate $\gamma/\wA$ by $n$
for the three plasma-current values, with indication of the three TAE families.}
\end{figure*}

To address this question, two additional magnetic equilibria are considered with
plasma currents $I_{-2\delta}$ and $I_{-5\delta}$ corresponding, respectively,
to reductions of size $2\delta$ and $5\delta$ of the reference value $\Iref$.
Their safety-factor profiles are plotted in \fref{fig:sensitivity.2} (left) and
$q_0$ now increases by $2\%$ and $5\%$ with respect to $\qref$. As a
consequence, the surface $q = 1$ is removed from the plasma and solutions of the
AE families $\aefamily{n}{l}{0}$ can exist only if $n < \frac{1}{2} (q_0 -
1)^{-1}$.

The new stability assessment is summarized in \fref{fig:sensitivity.2} (right
panel). According to predictions, AE families $\aefamily{n}{l}{0}$ are pushed to
lower $n$ and eventually vanish. For $I_{-2\delta}$ and before vanishing, AEs in
the family $\aefamily{n}{0}{0}$ have their growth rate reduced by $30\%$ when
compared to the reference case. The growth-rate reduction with respect to the
case $I_{-\delta}$ is even larger. However, the AE family $\aefamily{n}{0}{1}$
whose resonant surfaces are located at $q = 1 + \frac{3}{2 n}$ is also brought
to lower $n$ and inwards from its reference radial location.  For $I_{-5\delta}$
these AEs are located near the maximum gradient $\rmd n_\alpha/\rmd s$ and their
normalized growth rate peaks, accordingly, at $3.2\%$ for $n = 24$. In this way,
efforts to stabilize AEs by moving them out of the strong density-gradient
region are thwarted by the destabilization of AE families previously stable or
weakly unstable.

\section{Conclusions}

In summary, a perturbative hybrid ideal-MHD/drift-kinetic model was used to show that the
stability properties of ITER $\Ip = 15$~MA baseline scenario are significantly
sensitive to small changes of the safety-factor value on axis. Such small
variations were shown to cause large changes in the growth rate, toroidal
number, and radial location of the most unstable AEs. This sensitivity is not an
artificial feature of the ideal-MHD/drift-kinetic model employed to describe the
interaction between plasma species and AEs. On the contrary, it was shown to
proceed from the large value attained by the dimensionless parameter $\zeta$,
which is caused by the combination of large alpha-particle gyro-frequency [in
equation~\eref{eq:efficiency.condition}] with very low magnetic shear [in
equation~\eref{eq:linear.q}] throughout a substantial domain within the plasma
core.

If the large sensitivity of low magnetic-shear plasma configurations found in
this work is still present in nonlinear analysis, then detailed simulations
(e.g., suprathermal particle transport and redistribution by nonlinear
interaction with AEs) carried out in such circumstances should take this fact
into account, allowing a reasonable range of inputs in order to capture eventual
large changes in their results. Moreover, strong operational consequences should
also be expected, as such sensitivity would imply that AE instability and the
alpha-particle radial transport it entails are, in fact, unpredictable given the
extreme accuracy with which details about the safety factor profile would have
to be known near the magnetic axis.

\ack

This work was carried out within the framework of the EUROfusion Consortium and
received funding from the Euratom research and training programme 2014-2018
under grant agreement no.~633053. IST activities received financial support from
``Funda\c{c}\~{a}o para a Ci\^{e}ncia e Tecnologia'' (FCT) through project
UID/FIS/50010/2013. ITER is the Nuclear Facility INB no. 174. The views and
opinions expressed herein do not necessarily reflect those of the European
Commission, IST, CCFE, or the ITER Organization. All computations were carried
out using the HELIOS supercomputer system at the Computational Simulation Centre
of the International Fusion Energy Research Centre (IFERC-CSC) in Aomori, Japan,
under the Broader Approach collaboration between Euratom and Japan implemented
by Fusion for Energy and JAEA. PR was supported by EUROfusion Consortium grant
no.~WP14-FRF-IST/Rodrigues and NFL was supported by FCT grant no.~IF/00530/2013.

\section*{References}

\bibliographystyle{iopart-num}
\bibliography{papers.abb}

\end{document}